\documentclass[11pt]{article}
\usepackage{amsmath}
\usepackage{amsthm,amssymb}
\usepackage{color}
\usepackage{authblk}
\usepackage[numbers]{natbib}
\usepackage[T1]{fontenc}
\usepackage[utf8]{inputenc} 
\usepackage{subfig}
\usepackage{graphicx}

\usepackage{listings}
\date{}

\title{\Large \bf  Note on pre-taxation reported data by UK FTSE-listed companies. \\A search for Benford's laws compatibility }
\author[,1,2,3,4]{Marcel Ausloos \thanks{Corresponding author email: marcel.ausloos@uliege.be}}
\author[1,5]{Probowo  Erawan Sastroredjo }
\author[1,6]{Polina  Khrennikova  }

 \bigskip

\affil[1]{ School of Business, University of  Leicester, Brookfield, Leicester,   LE2 1RQ, UK;   email : (MA) ma683@leicester.ac.uk; (PES) pes13@leicester.ac.uk }   

\affil[2]{ Group of Researchers for Applications of Physics in Economy and Sociology (GRAPES), Rue de la belle jardini\`ere, 483,  B-4031 Angleur, Li\`ege, Belgium; email: marcel.ausloos@uliege.be} 

\affil[3]{ Department of Statistics, Predictions and Mathematics, Universitatea Babe{\c{s}}-Bolyai, Str. Mihail Kogălniceanu 1, 400084, Cluj-Napoca, Romania   }

 \affil[4]{ Department of Statistics and Econometrics, Bucharest University of Economic Studies, Calea Dorobantilor 15-17, 010552 Sector 1, Bucharest, Romania }

 \affil[5] { Department of Management, Parahyangan Catholic University, Bandung {\color{black}40141}, Indonesia  email: probowo27@gmail.com}
 
 \affil[6]{ Financial Engineering Group, Faculty of Behavioural, Management, and Social Sciences (BMS),
University of Twente, 7522 NB Enschede, The Netherlands;  e-mail: p.khrennikova@utwente.nl  }

\begin{document}
\maketitle  
\vskip -.5truecm
\newpage
\begin{abstract} Pre-taxation analysis plays a crucial role in ensuring the fairness of public revenue collection. It can also serve as a tool to reduce the risk of tax avoidance, one of the UK government's concerns. Our  report utilises pre-tax income ($PI$)  and total assets ($TA$) data from 567 companies listed on the FTSE All-Share index, gathered from the Refinitiv EIKON database, covering 14 years, i.e., the period from 2009 to 2022. We also derive the $PI/TA$ ratio, and distinguish between positive and negative $PI$ cases. We test the conformity of such data to  Benford's Law{\color{black}s},—specifically studying the first significant digit ($Fd$), the second significant digit ($Sd$), and the first and second significant digits ($FSd$). We use and justify two pertinent tests, the $\chi^2$ and the   Mean Absolute Deviation (MAD).  We find that both tests are not leading to conclusions in complete agreement with each other, - in particular the MAD test entirely rejects the Benford's Laws conformity {\color{black} of the reported financial data}. From the mere accounting point of view, we conclude that the findings {\color{black} not only cast some doubt on the reported financial data, but also}  suggest that many more investigations be envisaged on closely related matters.  {\color{black} On the other hand, the study of  a ratio, like $PI/TA$, of  variables which are (or not) Benford's Laws compliant add to the literature debating whether such indirect variables should (or not) be Benford's Laws compliant.}


\end{abstract}




\newpage
\section{Introduction}

{\color{black} Effective pre-audit accounting data analysis is crucial in detecting possitble earnings management and tax planning, thereby enhancing the expected integrity of financial reporting. In UK, the tax gap, i.e., the difference between the amount of tax that should theoretically be paid to HMRC, and what is actually paid, was estimated at £ 39.8 billion for the 2022 to 2023 tax year, accounting for  4.8\% of the known total liabilities

 (https://www.gov.uk/government/statistics/measuring-tax-gaps/1-tax-gaps-summary; accessed  on 29.01.2025).}

During our investigation of tax optimization by   listed FTSE companies, we had to use data on their pre-tax income ($PI$)  and data on their Total Assets ($TA$)  [or “$SIZE$” = $ln$ ($TA$)], in a given year, over a large time interval.   {\color{black}  In fact,} the ratio $PI/TA$ weighted through some irrelevant factor, at this stage, is a correction term to the deferred tax expense ($TXDI$) needed in order to calculate the (finally relevant) studied tax avoidance, estimated from the measure of total book-tax differences ($TBTD$) \cite{EMEC}. 

In so doing, we came across a huge data set for $TA$ and $PI$, i.e.,  6811 and 6768 data points, respectively.   {\color{black}  One pertinent question delves into the reliability of such pre-taxation data. Its analysis plays a crucial role in ensuring the fairness of public revenue collection. Econometrics, mixing statistical analysis and economic considerations, should increase   confidence to the population, and the more so to taxation officers.  The findings of this study, if the data is markedly found  bizarre,  can subsequently serve as a disciplinary tool toward reducing the risk of tax avoidance, and even tax evasion, both manipulations of profit shifts,  of great concerns to the UK government, - among others.
 
The Benford’s Laws, mathematically written in Section \ref{dataandmethodology},  state that the leading digits of (naturally occurring) numbers follow 
specific $log$ distributions \cite{Fewster2009}.   A deviation from this empirical distribution 
  is often considered as a warning suggesting a more detailed examination.}  Therefore, it has seemed to us of pertinent interest to observe whether such $PI$ and $TA$ data fulfils BLs expectations. This is the {\color{black} 
main} aim of this report{\color{black}, i.e.,  the research questions, 
     the distributions of the first-, second-, and first-second digit in such data are tested according to the empirical BLs, called BL1, BL2, and BL12 respectively. 
 
{\color{black} In other words, through the null hypothesis (agreement between the empirical data and expected laws)}, we aim to assess the  discrepancy between  secondary (financial) data and  their possible theoretical Benford distributions, through two statistical tests based on rather different concepts: the $\chi^2$ test and  the Mean Absolute Deviation (MAD), both outlined  in  Section \ref{findings}. 
 
Those BLs are {\color{black} known} of interest for  identifying 
irregularities in financial reports \cite{ref2Nigrini,Nigrini2012book,Kossovsky2014,ref3Miller,Mir2016,Morton2019,ref4Nigrini2017},   {\color{black} - but the literature is  too huge to be quoted here. Let us mention 
 that BLs have been considered in many fields, like finance, but also academia,  elections \cite{KukeliKarunaratne20}, engineering, medicine, psychology, physics, religion, scientometrics, sport, and likely many others; some of these were 
  discussed in  \cite{EMEC}, on which the following subsection sometimes overlaps.} 

Moreover, because of the  $PI$ and $TA$ data type and size, it is possible to distinguish between negative and positive $PI$s, whence to study them in the BLs framework. Notice that this sign distinction has rarely been considered; indeed most authors, except a few to our knowledge, are searching for the BL obedience in absolute values of data \cite{ref4Nigrini2017}.

{\color{black} A short and focused literature review is found in Section \ref{LitRev}. Such a literature is huge. In order not to add to useless (and, necessarily or often, incomplete) literature reviews, we reduce the present literature review section to the essential papers pertinent to our aim; i.e., we include a mention of  (i) the pioneering papers, (ii) the next most often quoted ones, whence likely of general interest and (iii) the most modern ones, essentially to pin point the “state of the art”.  We focus on papers at  the intersection of 3 sets: (i) high order BLs, - in particular BL2 and BL12, (ii) financial data relevant to pre-taxation, (iii) statistical tests,-  in particular papers considering the Mean Absolute Deviation (MAD).}
 
A warning: irregularities may not always be revealed through the use of Benford's Law \cite{Druica et al. 2018,Pavlovicetal2019,CerquetiMaggi2021b,JianuJianu21}. Moreover, financial data   might  not necessarily be strictly conforming to BLs: it may depend on the data ranges \cite{Ausloos et al. 2021,Herteliiuetal2021}.

Notice that within a statistical framework, it is sometimes debated whether BLs compliance can be extended to derived, correlated, or combined quantities. Therefore, we also calculated $PI/TA$ when possible, i.e., when both $PI$ and $TA$  data are reported  for a company in a given year.
 
To conclude these remarks, a fundamental limitation of  studies reported so far should be pointed out: it is their common focus on  BL1 tests , - except for a few authors who have approached income items and BL2 from a behavioural perspective \cite{NigriniMiller2009,Ileanuetal2019,Coracioni2020,Osieckietal2024}. Consequently, our research aims to significantly advance the literature by conducting an in-depth statistical analysis of  (accounting and tax) variables in financial statements, exploring potential conformity to BL1, BL2, and BL12.

We fully  present the data acquisition and study methodology in Section \ref{dataandmethodology}. {\color{black} We run two null hypothesis significance testing methods to investigate conformity to BLs.}
More precisely, as measures to assess the  discrepancy between the empirical and the theoretical Benford distributions, we use the $\chi^2$ test and  the MAD test  as described in detail in  Section \ref{findings}. We report our findings through Tables and Figures.   {\color{black}  Interestingly, we find that the MAD test entirely rejects the Benford's Laws conformity of the reported financial data, whence the relevant null hypothesis}.

We conclude in Section \ref{conclusions}, both with remarks on the statistical tests and their {\color{black} disagreement}, and on the implications for practical accounting  research.  
  {\color{black}   Indeed, we find that both $\chi^2$ test and MAD tests are not leading to conclusions in complete agreement with each other.  Thereby, one rejects the null hypothesis. Therefore, this finding demands further studies on the validity ranges of statistical tests applications. Notice that the study of  a ratio, like $PI/TA$, with variables, - which are (or not) Benford's Laws compliant, as {\it a posteriori} differently found, add to the literature debating whether indirectly measured  variables should be (or not) Benford's Laws compliant. }   
  
  {\color{black}  Furthermore,  from the mere accounting point of view, on one hand,  we conclude that the findings not only cast some doubt on the reported financial data, but, on the other hand, thereby suggest that many more empirical and thorough investigations be envisaged on closely related financial data so reported by listed companies. }
}

\section{Literature Review} \label{LitRev}

 {\color{black}The Benford's laws have been like a “sleeping beauty” sleeping in the dirty pages of logarithmic tables \cite{JASISTmirbeauty}.  
But it has been revived in strategic management literature \cite{Fahimifaretal24} and in accounting \cite{ref4Nigrini2017,ThomasAR89}, - fields of interest here. The range of applications is huge, i.e., as long as there are large natural data sets.  For conciseness, we focus on papers at  the intersection of 3 sets: (i) high order BLs, (ii)   tax concerned financial data, (iii) specific statistical tests, in particular papers considering the Mean Absolute Deviation (MAD). We also comment on pertinent papers at the intersection of the pair of such sets.}

 {\color{black}In brief, one should start recognizing the pioneer work of Nigrini \cite{ref2Nigrini,Nigrini2012book,ref4Nigrini2017,NigriniIEEE1999}: he introduced   and 
  developed  statistical research  based on BLs  to estimate compliance of taxpayers, and  of companies engaging in tax planning strategies.
In \cite{ref4Nigrini2017}, Nigrini provides a review of the literature on audit sampling and   perspectives. }

 {\color{black}
One of the most relevant papers at the here above mentioned  triple set intersection is that of Alali and  Romero in  2013  \cite{Alali and Romero 2013}.
 They discovered significant differences in various accounting figures within US financial statements of companies, - either audited by Big Four firms or by non-Big Four firms. Thereafter, Druica et al. studied Romanian banks along BL2  \cite{Druica et al. 2018}.  Also, Prachyl and  Fischer evaluated the conformity of municipality financial data with BL2  \cite{PrachylFischer20}. Cheuk et al. assessed the  "financial reporting quality of Company Limited by Guarantee charities in Malaysia" \cite{Cheuketal21}. 
 
 More oriented toward comparing tests, Kössler et al. looked at share prices \cite{Kossleretal24} essentially through BL2.  In the same line of thoughts, da Silva Azevedo  \cite{daSilvaAzevedo21} as well as  Cerqueti and Lupi \cite{CerquetiLupi2023} considered BL12.

 In a more comprehensive way, Sadaf \cite{Sadaf17}, Patel et al.  \cite{Pateletal22}, as well as Sylwestrzak  \cite{Sylwestrzak23}, studied both BL2 and BL12 on various possible data manipulation by managers.}
 
 {\color{black}
Concerning the pertinent literature at the intersection of pairs of sets, let us  recognize the keystone and pioneering observation of Carslaw in 1988 \cite{Carslaw88}: he  observed, in fact working
on BL2,  that New Zealand companies  income statements showed a markedly higher occurrence of 
0s and a lower occurrence of 9s  in the second digit position than should be expected, - thereby implying voluntary roundings. Similarly findings were obtained by Niskanen and Keloharju in 2000 on Finnish public companies  \cite{NiskanenKeloharju2000}. 
Using both BL2 and BL12, Ausloos et al. also found that "Benford’s laws tests on S\&P500 daily closing values and the corresponding daily log-returns both pointed to huge non-conformity" \cite{Ausloos et al. 2021}.  Similar  studies on two joint BLs by Das et al. \cite{Dasetal17} and by Jordan and Clark \cite{JordanClark22}  can be mentioned.

 For completeness, let us also mention that  authors have used various values of  financial
results  and tests 
 to detect  potential data manipulation. Of values with respect to our report are, e.g.,
Van Caneghem \cite{VanCaneghemEAR04} investigated with the aid of BL2 tests a sample of 1256 UK companies that reported pre-tax income for the accounting year 1998. He found results similar to those of Carslaw \cite{Carslaw88}.  Other recent works  implying BL2 studies are in  \cite{Jordan et al 2009},    \cite{Istrate2019}, 
\cite{GrammatikosPapanikolaouJFSR21}, \cite{RanadeGandhi2022}, \cite{HarbetalJAAR23}, \cite{IstrateCarpEEE24}. Last but not least,   Günnel and Tödter considered BL12 \cite{GunnelTodterEmpir09}, like Le and Mantelaers who even discussed the state of the art up to BL123 \cite{LeMantelaersMAB2024}, or Sardar and Sharma studying BL3 and BL4  on financial reports of  several listed companies of the Adani Group \cite{SardarSharmaAdani}.

  Of course, many  other works discuss statistical tests and financial reports sometimes using $\chi^2$ or MAD tests;  many other statistical tests are available and are considered. However,  because such papers restrict their consideration to BL1,  for the sake of this literature review finiteness, we  repeat that we limit ourselves to the above works for framing our field of interest.
  }

\section{Data Acquisition and Study Methodology} \label{dataandmethodology}
 
\subsection{Data Acquisition}
We utilise data from Refinitiv EIKON, focusing on pre-tax income ($PI$) and the total assets ($TA$) as pre-taxation indicators from 2014 to 2022, both years being included in the time interval  for the study. The financial data   is  recorded in GBP currency, ensuring uniformity in currency representation across all datasets. The FTSE All-Share list is updated annually. 

We  should emphasise several steps throughout the data retrieval from the Refinitiv EIKON database. Firstly, we utilise the list of companies indexed in the FTSE All-Share   the (final)  year 2022, which comprises 567 companies, - because 2022  marks the commencement of our investigation into tax avoidance \cite{EMEC}. Secondly, we have encountered a lot of missing data. To maintain the authenticity of the dataset, we leave these data points blank in our "Master Data Bank". 
 Lastly, to facilitate our calculations, we employ the term "Absolute Fiscal Year (FY)" within Refinitiv EIKON, which defines the financial period as a continuous 12-month interval commencing on January 1 and concluding on December 31. This warning is emphasized since the "real" UK fiscal year usually goes from April in a given year  till March the next year.
 
 The data extraction process Data Extraction Process
from the Refinitiv Eikon database is based on a connection to a server at the University of Leicester. We first download and installed the file using the link $https://eikon.refinitiv.com
$ to install the Eikon add-in for Excel. It is important to note that this add-in is exclusively compatible with non-Mac computers. Upon completion of the download, we launch Microsoft Excel. Within the "File" menu, we select "Options" to enable the Refinitiv Eikon add-in. After making the required adjustments, it is advisable to restart Excel by closing and reopening the application. Following this action, the Refinitiv Eikon should be successfully integrated with Microsoft Excel.

After entering the instrument, we searched for "total assets" and “pre-tax income”. We used "total assets reported," representing the company's total asset;  we used the net income before taxes to describe pre-tax income representing the sum of operating income. Many other financial data are available on Refinitiv Eikon database, but are not presently discussed here.


In so doing, one obtains  6768 and 6811 data points for $PI$ and $TA$ respectively. Due to the large size of the data set, one can split the $PI$ data into $PI^{(-)}$	 and $PI^{(+)}$, and analyze them independently, in order to observe if any compliance holds for both types of pre-tax incomes. 

We wish to warn the reader that a few data points  are markedly (i.e., after displaying the pertinent histograms, logically deduced to be) incorrect, but this can be assumed to be resulting from a transcription error of the decimal point. As easily understood, this typo is irrelevant for a BLs study.  However, a small number  ($\simeq 3$) of data points reported by Refinitiv EIKON are readily ''anomalously extreme outliers'', - with quite incomprehensible values.  We do not have access to the sources used by Refinitiv EIKON and cannot pursue further the "correction" of these data points. Rather than, as often done, winsorising whence modifying the first few digits of these numbers, we have preferred conserving  them.  This voluntary step, amounting to not disregarding 3 bizarre data points, seems of very weak influence for the BLs analysis due to the   large number of data points, -  called $N$ in Table \ref{tablePITAPI/TAstats}.

 \begin{figure}
\centering
\centering {\includegraphics[height=15cm,width=15cm] {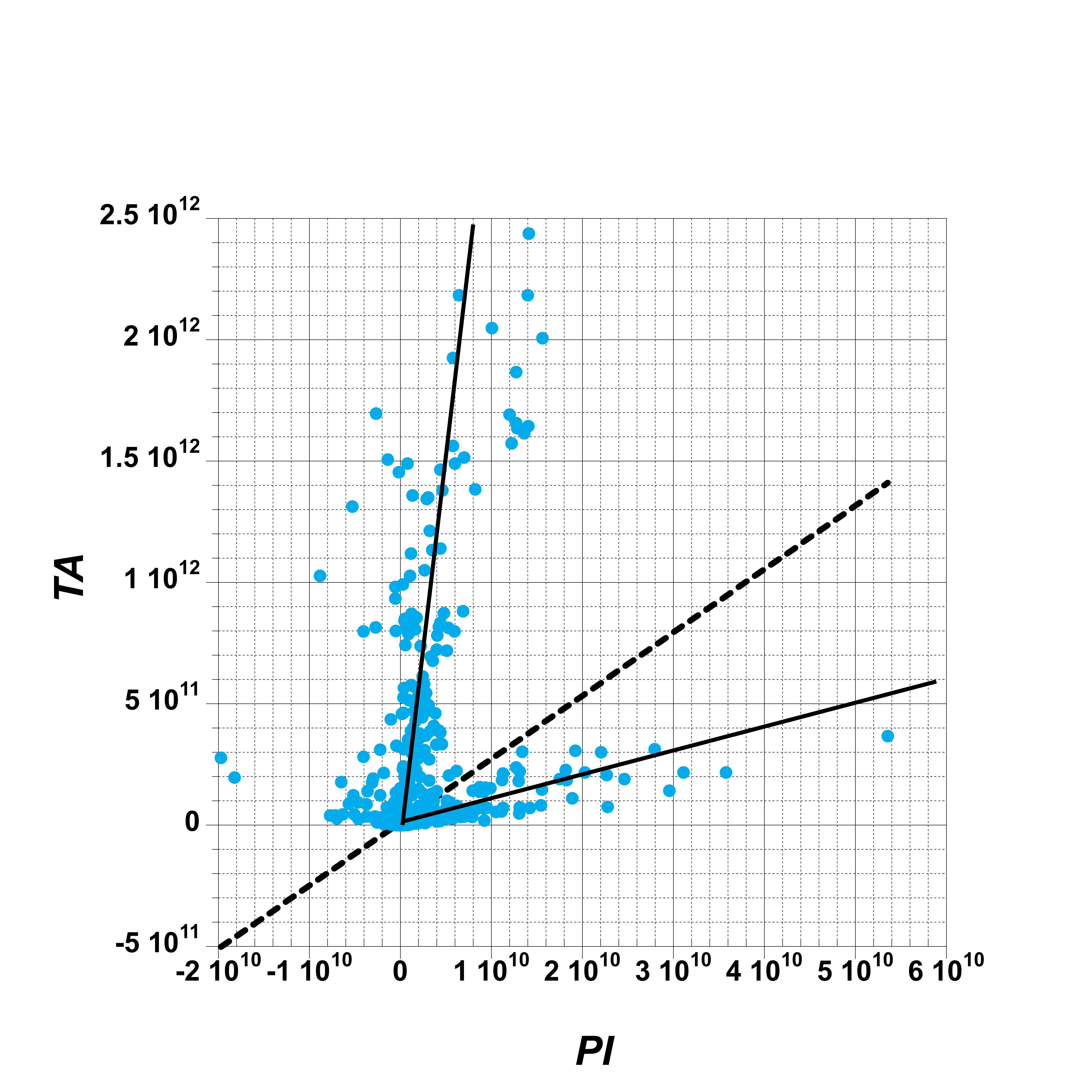}}
\hfill
\caption{Display of the "empirical relationship(s)" between   $TA$ and  $PI$, - when both exist,  for the listed FTSE companies between 2009 to 2022; both straight lines may point to 2 possible types of companies "tax strategies"; the dash line is the "official" linear regression line; the full lines are guides to the eye only.} \label{figPlot4TAPI}
\end{figure}

A "detail" of financial interest appears when observing whether a relationship might exist between $TA$ and $PI$. It is displayed on Fig. \ref{figPlot4TAPI}. "Interestingly", it seems that there are  two clusters. Whether this implies to deduce different pre-tax income strategies by companies seems to be a theoretical question in itself and demands some thought before further modelling, but the matter is much outside the present Note and is left for further investigation.

As justified in the Introduction, we calculated $PI/TA$ for common (company-year) cases before further analysis; one gets 6765 data points, - due to the absence of one or the other data for a given company in a given year, - in particular the 3 "anomalous outliers".

The statistical characteristics are given in Table \ref{tablePITAPI/TAstats}.

  \begin{table}\begin{center}  \fontsize{8pt}{8pt}\selectfont
   \caption{  Summary of  the (rounded) main statistical characteristics  of the downloaded $PI$, further  distinguishing between $PI^{(-)}$	 and $PI^{(+)}$,  and $TA$, and the subsequently deduced $PI/TA$.    {\color{black}  Financial data are originally  in £}.} \label{tablePITAPI/TAstats}
\begin{tabular}{|c||c||c|c|c|c|c|c|c|c|} 
 \hline 
 	&	$N$&	min.			& Max. 			&    Mean    		&   St. Dev.   	&   Skew.    	&   Kurt.    		& CV 	 \\ \hline \hline
 $PI$	&	6768	&	-19722854000	&	53579321000	&	316593080 	&1752653700	& 11.26074	&	227.834	&5.53598	\\
$TA$	&	6811&	15580 		&     2438029900000&	20513747000 	&130724180000& 10.51739 	&	128.530 	&6.37252 \\ \hline
\hline
$PI^{(-)}$&	1160	&	-9.9748	&	-1.0000	&	-3.7738	&	2.4381	&	-0.79236	&	-0.50234	&	-0.6461	\\
$PI^{(+)}$&	5608	&	1.0000	&	9.9996	&	3.8908	&	2.4777	&	0.78672	&	-0.52936	&	0.6368	 \\ 
\hline
 \hline
$PI/TA$	&	6765	&	-1.81579	&	3.11173	&	0.07390	&	0.16680	&	4.14856	&	82.5632	&	2.25731\\ 
$PI^{(-)}/TA$&	1158	&	-1.81579	&	-0.00004	&	-0.10559	&	0.16054	&	-4.33229	&	27.73320	&	-1.52035	\\
$PI^{(+)}/TA$&	5607	&	0.00002	&	3.11173	&	0.11096	&	0.14221	&	9.97602	&	161.7935	&	1.28154 \\ \hline
\end{tabular}
 \end{center}
 \end{table}

\subsection{Study Methodology}\label{methodology}

The study of pre-income tax (and total assess) reported data by FTSE-listed companies as being  BLs compatible start from
  the following formulae for the first digit, (BL1), i.e., for calculating the probability that a   number will have a first digit as non zero \cite{Ausloos et al. 2021}:
\begin{equation}
	P(d_1=i)=log_{10} ( 1+\frac{1}{i})	 \;\; ,
	\end{equation} \label{BL1}
for $ i$ = 1, 2, 3, 4, 5, … 9. So, $P(d_1=i)$  represents the likelihood that  any number starts with a digit ($i$), between 1 and 9, according to a $log$ law. 
In other words, according to the first digit Benford's Law (BL1), any number from the selected data set should  start with “1”,   (approximately 30.1\% of the time;   “9” as the least frequently occurring first digit should approximately 4.6\% of the time as the first digit.

 This formula BL1 can be extended to the first two digits, resulting in a BL12 \cite{Ausloos et al. 2021}:
\begin{equation}
	P_{12} (d_1 \;d_2\;=\;i\; j)=log_{10} ( 1+ \frac{1}{d_1 \;d_2}	).
	\end{equation} \label{BL12}
In this context,  $d_1$ indicates the first digit ($i$), $\in$ [1, 9], while $d_2$ signifies the second digit   {\color{black}  ($j$) of a number}, which maybe a 0. Thus, $d_1 \;d_2\; \in$ [10; 99]. 

 Furthermore, a Benford's ($log$) Law can also be applied to test the empirical occurrence of the second digit in the data set numbers, using the following formula,  called BL2 \cite{Ausloos et al. 2021}:
\begin{equation}
	P(d_2=j)= \sum_{k=1}^9\; log_{10}  ( 1+\frac{1} {10 \;k + j }  ) \; \; ,
	\end{equation} \label{BL2}
for $j$ = 0, 1, 2, 3, 4, … 9 and $k$ = 1, 2, 3, … 9. So, $P(d_2=j) $ represents the likelihood that the second digit ($j$) of a number is any digit between 0 and 9. The calculation methodology is relatively straightforward: $k$ ranges from 1 to 9; consequently, 10 $k$ serves as a multiple of these integers, resulting in the series of numbers 10, 20, ..., and 90.  
 
 For further readers' easiness, we report the expected frequency values of the pertinent digits in BL1, BL2, and a few for BL12 in Table \ref{tableBL1BL2BL12}.
 
  {\color{black}  One may wonder about the origin and significance of such BLs, - a "a mysterious law of nature" \cite{Whymanetal2016a,Whymanetal2016b}. The matter is still an open question. Roughly speaking, a $y=log(x)$ law can be contrasted to the $y=exp(x)$ law. In the latter case, it results from the "growth equation" $dy/y=dx$, leading to the Keynes infinite growth, e.g., in economy. In contrast, the former arises form $dy=dx/x$. This means that an elementary  increase $dy$  has a reducing effect (by $1/x$)  due to the accumulation of previous $x$s. In economy, it corresponds to one of the Marxist principles. The BL1, as the basic example,  derives from
  \begin{equation}\label{BLeq4}
  \frac{dy}{dx} \;=\; -\;\frac{1}{x(x+1)}\; 
  \end{equation}
  meaning a reduction of previous values, whence subsequent $log$ decay, through a complex factor $-1/[x(x+1)]$. Notice that such a factor occurs in calculating the geometric means (of size $x$ quantities) \cite{Whyman21}.
  }
  
  \begin{table}\begin{center}  \fontsize{10pt}{10pt}\selectfont
  \caption{Expected proportion values for BL1, BL2, BL12.} \label{tableBL1BL2BL12}
\begin{tabular}{|c||c|c||c|c|c|c|} 
 \hline 
 $d$	& BL1	& BL2 & BL12 & \textbf{$d_1 d_2$}\\
  \hline	\hline			
0	&		&0.1197	&		& \\	
1	&0.3010	&0.1139	&0.0414	&10\\
2	&0.1761	&0.1088	&0.0378	&11\\
3	&0.1250	&0.1043	&0.0348	&12\\
4	&0.0969	&0.1003	&0.0322	&13\\
5	&0.0792	&0.0967	&0.0230	&14\\
6	&0.0669	&0.0934	&0.0280	&15\\
7	&0.0580	&0.0904	&0.0264	&16\\
8	&0.0512	&0.0876	&0.0248	&17\\
9	&0.0460	&0.0850	&0.0235	&18\\
 	& 		& 		&  ...		&  ...\\ 	
	& 		& 		&  0.0086	&50\\ 	
	& 		& 		&  ...		&  ...\\
	&		&		&0.0044	&99\\    \hline				
\end{tabular}
 \end{center}
 \end{table}
 
\section{Findings} \label{findings}

We report the results in the following Tables:
\begin{itemize}
\item    
Table \ref{tablePIBL1} countains the (rounded) First digit ($Fd$)  of the downloaded   $PI$, as well as $PI^{(-)}$ and $PI^{(+)}$,  and their pertinent frequencies to be compared to the BL1. {\color{black} From the results of the two tests, we can notice that the $PI$ variable shows non-conformity (both in the ``gain" and ``loss"  domains) considering the MAD test. The $\chi^2$  analysis shows mixed results pointing to some violation for the $PI$ variable and its ``gain" domain. Notably, we observe ``under-occurrence" of the last 3 digits (7,8,9) as the first digit concerning BL1 distribution, potentially pointing to corporate ``rounding up" practices. }

 \item
 Table \ref{tableTAPITABL1} contains the (rounded) First digit ($Fd$)    occurrences of the downloaded  $TA$ and the deduced $PI/TA$ values and their pertinent frequencies to be compared to the empirical BL1. {\color{black} For the total assets and the  $PI/TA$ ratio, we again observe a complete non-conformity based on the MAD test and non-conformity of the ratio variable considering the $\chi^2$ analysis. We can observe for the $TA$ variable an ``over-occurrence" of the last digit $(9)$, which also leads to some interesting behavior for the occurrence of $(1)$ and $(9)$  as  in the first digit of the $PI/TA$ ratio. The findings can be of interest for suggesting a deeper investigation of different ratios as part of financial statement analyses. }
 
 \item   
Table \ref{tablePIBL2} reports the (rounded) Second digit ($Sd$)  occurrence of   $PI$  also distinguishing between $PI^{(-)}$	 and $PI^{(+)}$,  and the pertinent frequencies for comparison to BL2. {\color{black} We can observe that the results of the BL2 analysis are more mixed, with an absolute non-conformity for both ``loss" and ``gain" domains of pre-tax income with the MAD test,  and a conformity of all second digits, considering the $\chi^2$ analysis. } 

\item      
  Table \ref{tableTAPITABL2} summarizes the (rounded) Second digit ($Sd$)   occurrence of  $TA$ and $PI/TA$    and their pertinent frequencies to be compared to the BL2. {\color{black} For the $TA$ variable and the derived ratio, we observe the same picture as for the $PI$ analysis: non-conformity with MAD for all variables and also non-conformity considering the behavior of the $PI/TA$ ratio with the $\chi^2$ test. For the ratio variable, we can observe an ``over-occurrence" of the $(0)$ and $(1)$ as well as $(9)$ as  second digit. These observations point to possible irregularities in $PI$ reporting and in the valuation of companies' assets. } 
  
 \item
 Table \ref{tablePIBL12} reports the (rounded)  First- and Second-  digit ($FSd$) of the downloaded  $PI$,  again further distinguishing between $PI^{(-)}$	 and $PI^{(+)}$,  and the pertinent frequencies to be compared to   BL12.
 
  \item 
Table \ref{tableTAPITABL12} contains a summary of the (rounded) First- and Second digit ($FSd$)  of   $TA$ and $PI/TA$ values and their pertinent frequencies to be compared to the BL12.   

{\color{black} Based on  results in Table \ref{tablePIBL12} and  Table \ref{tableTAPITABL12}, the BL12  conformity leads to a relatively complex analysis, especially considering the evaluation of derived variables (such as the $PI/TA$ ratio). The results from Tables \ref{tablePIBL12}-\ref{tableTAPITABL12} are again characterized by non-conformity when considering the MAD test. But the $\chi^2$ statistics point to the conformity of the $PI$  values in the ``gain" domain. Potentially, as a result, this leads to non-conformity of the $PI/TA$ ratio. } 
 \end{itemize}

\begin{figure}
\centering
\centering {\includegraphics[height=15cm,width=15cm]  {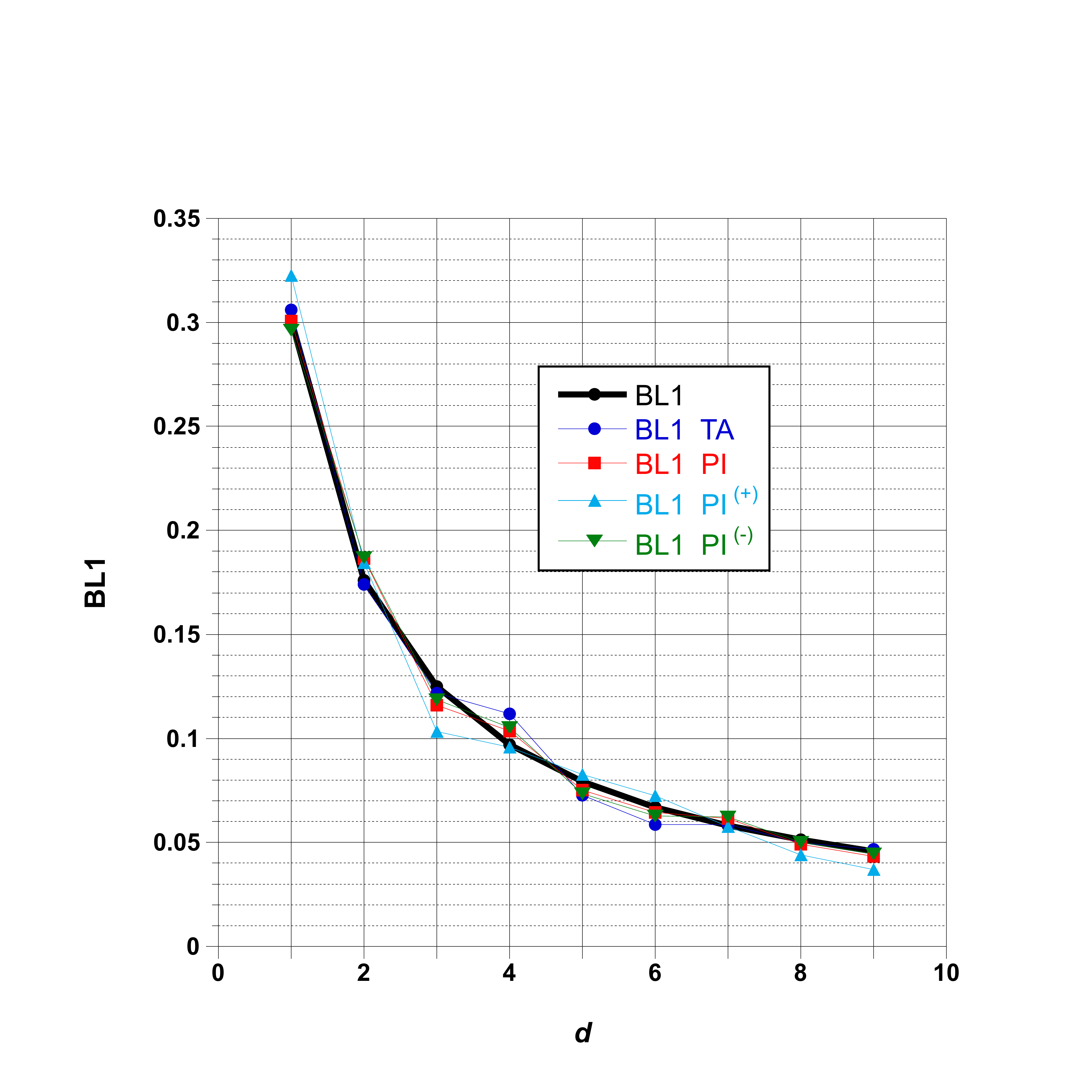}}
\hfill
\caption{Display of BL1 data analysis for $TA$,  $PI$, $PI^{(+)}$, and $PI^{(-)}$ of listed FTSE companies between 2009 to 2022; the empirical BL1 is also shown. }  \label{figPlot18BL1}
\end{figure}

\begin{figure}
\centering
\centering {\includegraphics[height=15cm,width=15cm] {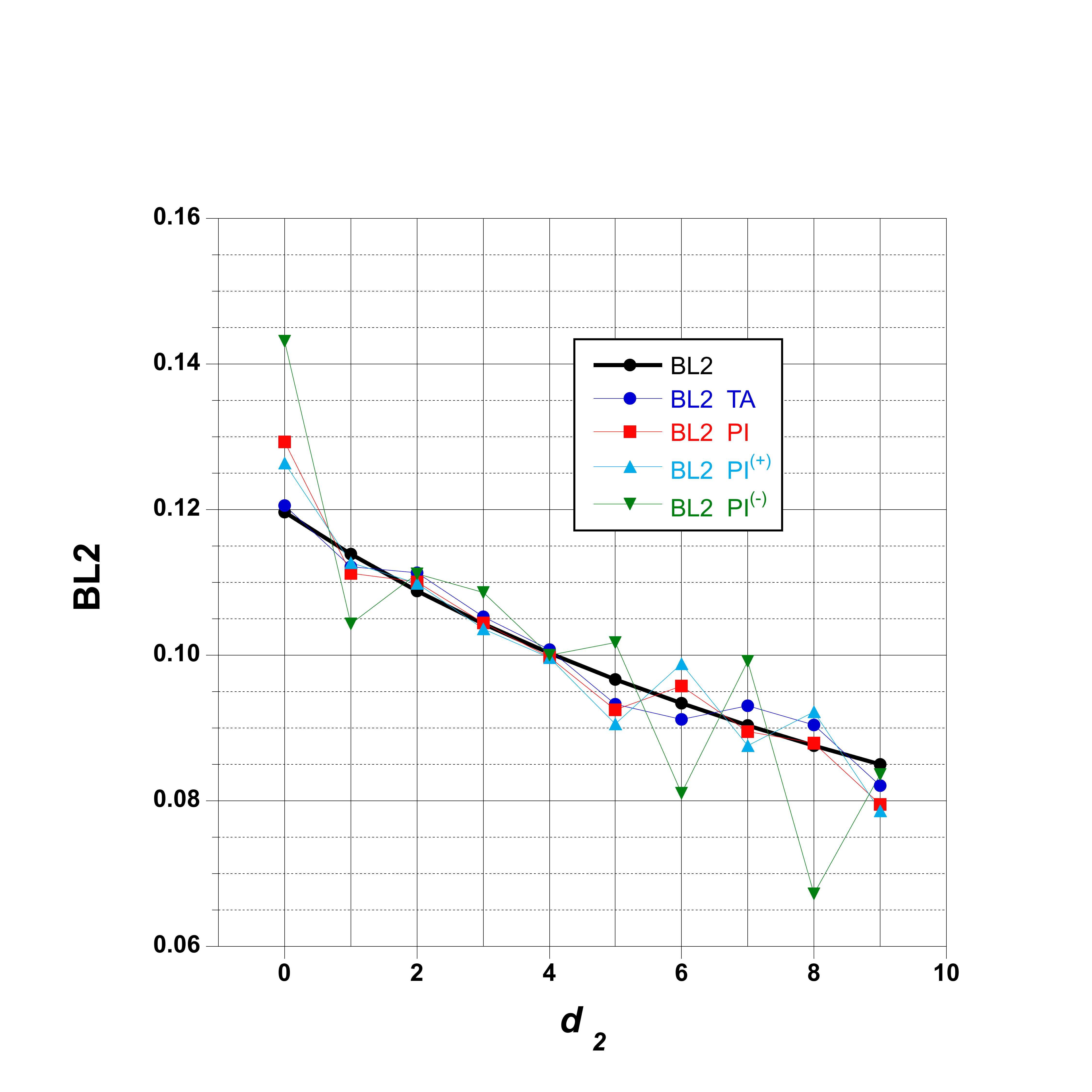}}
\hfill
\caption{Display of BL2 data analysis for $TA$,  $PI$, $PI^{(+)}$, and $PI^{(-)}$ of listed FTSE companies between 2009 to 2022; the empirical BL2 is also shown. } \label{figPlot19BL2}
\end{figure}

\begin{figure}
\centering
\centering {\includegraphics[height=15cm,width=15cm] {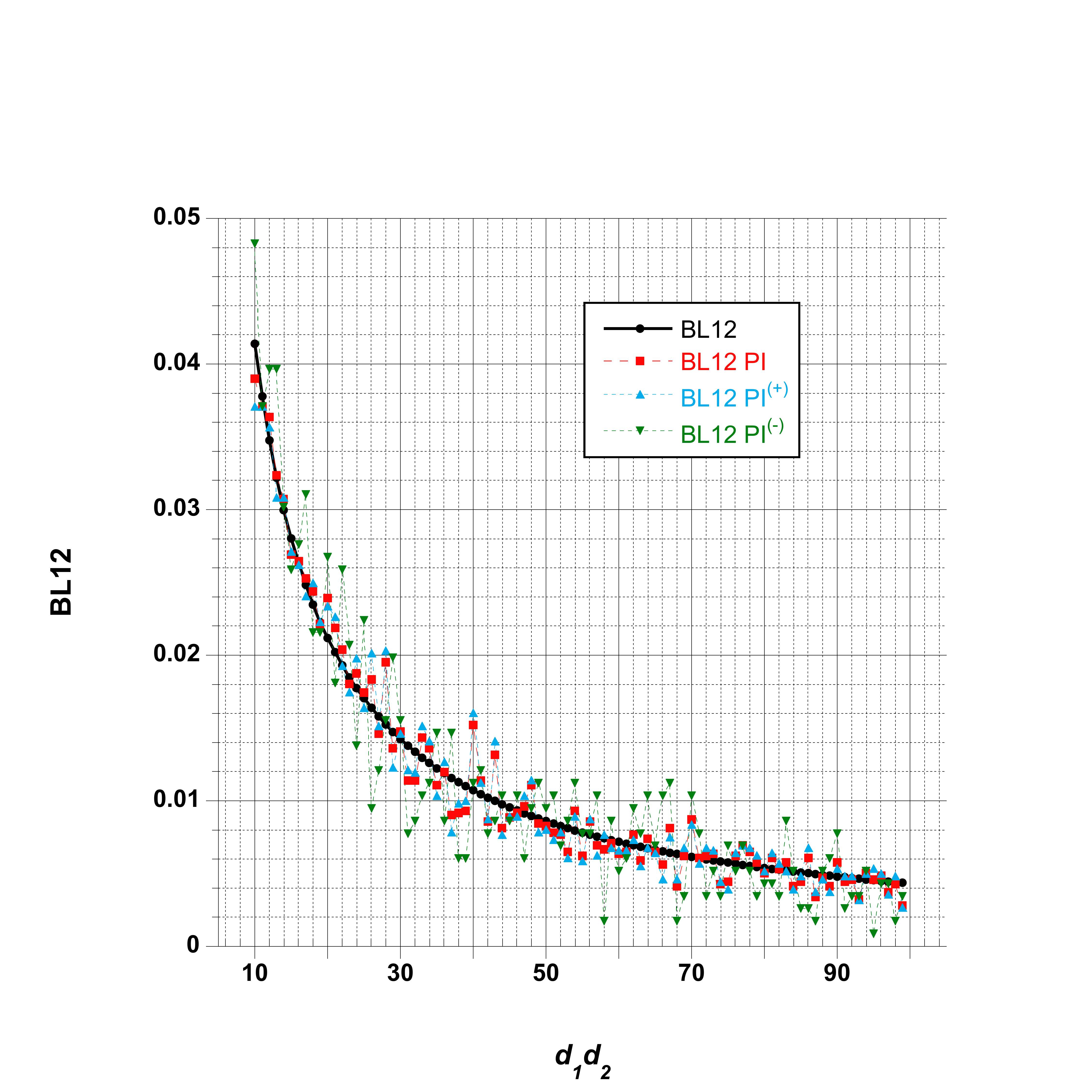}}
\hfill
\caption{Display of BL12  data analysis for  $PI$, $PI^{(+)}$, and $PI^{(-)}$ of listed FTSE companies between 2009 to 2022; the empirical BL12 is also shown.}  \label{figPlot12BL12}
\end{figure}

The analysis of BL1 and BL2 is also illustrated in Fig. \ref{figPlot18BL1}. and \ref{figPlot19BL2} respectively. The BL12 is  illustrated on Fig. \ref{figPlot12BL12}, - without the $TA$ data, - otherwise the figure is somewhat unreadable. {\color{black}Let it be observed that supplementing the data reported in  the Table \ref{tablePIBL12} and  Table \ref{tableTAPITABL12}  with that in Fig. \ref{figPlot12BL12},  some salient irregularities appear in BL12,  in particular in the last higher two digits' occurrence (such as $60, ..., 99$). The  worst is for the $PI^{(-)}$ data which seems to have much (unexpected) deviation from the empirical BL12. }

The results from the statistical tests, in order to assess the  discrepancy between the  observed  frequency distributions and the theoretical BL distributions, are reported in Tables \ref{tablePIBL1} -   \ref{tableTAPITABL12}. Recall that the $\chi^2$ test and  the MAD tests concern two different concepts of distances: 
\begin{itemize}
\item the $\chi^2$, obviously as usual defined through the number of observations ($O$) and the number of expected ($E$) ones, distributed in a number of bins equal to $D+1$, where $D$ is the number of degrees of freedom,   
\begin{equation}\label{chi2eq}
\chi^2 \; =\;  \sum_{i=1}^{D+1} \; \frac{(O_i-E_i)^2}{E_i}
\end{equation}
 is classically used, but is said to tend toward rejecting Benford compliance of observations even when the deviations from the theoretical BL ($E$) are negligible, — mainly in  large samples.
\item the MAD, defined through the observed ($f_o$) and the expected ($f_e$) frequencies, in the pertinent number $K$ of bins, as 
\begin{equation}\label{MADeq}
MAD \; = \;  \sum^K \; |\; f_o \; - \; f_e\; |
\end{equation}
 is  thought to be the most reliable test for checking the validity of the BL \cite{CerquetiLupi2023}, - but not always \cite{Druica et al. 2018}. In brief, for BL1,  a value below 0.006 allows to deduce  close conformity, while a MAD between 0.006 and 0.012 refers to only acceptable conformity; marginally acceptable conformity occurs for values between 0.012 and 0.015; nonconformity  is the conclusion otherwise. For BL2, close conformity occurs for MAD $\leq$ 0.008, while close conformity occurs for MAD $\leq$ 0.0012 for BL12 \cite{ref2Nigrini}.
\end{itemize}

  {\color{black}  Of course, many different statistical tests have been used for assessing the conformity (of financial data, for example) to BLs: a short list may include, without any hierarchical order, the Kolmogorov–Smirnov test, the Chebyshev distance, the Kullback–Leibler divergence,  the Freedman-Watson ($U^2$)  test, the Joenssen JP-square test,  the $z$-statistics, the Financial Statement Divergence Score, the Kuiper test; the Binomial Probability test, and even the Euclidean distance or  regression approaches and other "smooth tests". 
  
  Recent discussions and works of interest are covered with pertinent recommendations  by Cerqueti and Maggi \cite{CerquetiMaggi2021b},     Lesperance et al. \cite{Lesperanceetal16}, Ducharme et al. \cite{Ducharmeetalarxiv}, Henselmann et al.   \cite{Henselmann et al. 2015}, 
Cerqueti and Lupi \cite{CerquetiLupi2023,CerquetiLupi2021}, and  Barabesi et al. \cite{Barabesietal21}.}


   \begin{table}\begin{center}  \fontsize{8pt}{10pt}\selectfont
   \caption{  Summary of  the (rounded) First digit ($Fd$)  of the downloaded  $PI$, further  distinguishing between $PI^{(-)}$	 and $PI^{(+)}$,  and the pertinent frequencies to be compared to the BL1.} \label{tablePIBL1}
\begin{tabular}{|c||c|c|c||c|c|c||c|c|c|} 
 \hline 
$d_1$ &$Fd$ $ PI$&$Fd$  $PI^{(-)}$& $Fd$  $PI^{(+)}$&   BL1 $PI$& BL1   $PI^{(-)}$ &   BL1 $PI^{(+)}$  & BL1  \\\hline 
1	&	2035	&	374	&	1661	&	0.30103	&	0.30068	&	0.32241	&	0.29618	\\	
2	&	1262	&	214	&	1048	&	0.17609	&	0.18647	&	0.18448	&	0.18688	\\	
3	&	785	&	120	&	665	&	0.12494	&	0.11599	&	0.10345	&	0.11858	\\	
4	&	701	&	111	&	590	&	0.09691	&	0.10358	&	0.09569	&	0.10521	\\	
5	&	508	&	96	&	412	&	0.07918	&	0.07506	&	0.08276	&	0.07347	\\	
6	&	436	&	84	&	352	&	0.06695	&	0.06442	&	0.07241	&	0.06277	\\	
7	&	416	&	67	&	349	&	0.05799	&	0.06147	&	0.05776	&	0.06223	\\	
8	&	332	&	51	&	281	&	0.05115	&	0.04905	&	0.04397	&	0.05011	\\	
9	&	293	&	43	&	250	&	0.04576	&	0.04329	&	0.03707	&	0.04458	\\	
\hline \hline
$\chi^2$ &16.5706&10.3233&15.7448&&&&\\
$MAD$ &&&&0.04103&0.07764&0.04664&\\
	\hline
	 &\multicolumn{3}{|c||}{$\chi^2_c$ =   15.507, at  5\%,\; for D = 8. } &\multicolumn{3}{|c||}{close conformity if MAD $\leq$ 0.006
	 }  & \\	\hline
	\end{tabular}
 \end{center}
 \end{table} 

   \begin{table}\begin{center}  \fontsize{8pt}{10pt}\selectfont
   \caption{  Summary of  the (rounded) First digit ($Fd$)  counted occurrences of the downloaded  $TA$ and $PI/TA$ values and their pertinent frequencies to be compared to the empirical BL1.} \label{tableTAPITABL1}
\begin{tabular}{|c||c|c||c|c||c|} 
 \hline 
$d_1$ &$Fd$ $ TA$&$Fd$  $PI/TA$&     BL1 $TA$& BL1   $PI/TA$   & BL1 \\\hline 
1	&	2084	&	2223	&	0.30598	&	0.32860	&	0.30103	\\
2	&	1186	&	1100	&	0.17413	&	0.16260	&	0.17609	\\
3	&	829	&	655	&	0.12171	&	0.09682	&	0.12494	\\
4	&	761	&	564	&	0.11173	&	0.08337	&	0.09691	\\
5	&	494	&	523	&	0.07253	&	0.07731	&	0.07918	\\
6	&	400	&	508	&	0.05873	&	0.07509	&	0.06695	\\
7	&	399	&	446	&	0.05858	&	0.06593	&	0.05799	\\
8	&	340	&	386	&	0.04992	&	0.05706	&	0.05115	\\
9	&	318	&	360	&	0.04669	&	0.05322	&	0.04576	\\	\hline \hline
$\chi^2$ &27.757&117.287&&&\\
$MAD$ &&&0.04258&0.11404&\\
	\hline
	 &\multicolumn{2}{|c||}{$\chi^2_c$ =   15.507,  at  5\%,\; for D = 8. } &\multicolumn{2}{|c||}{close conformity if MAD $\leq$ 0.006}  & \\	\hline
	\end{tabular}
 \end{center}
 \end{table} 

 \begin{table}\begin{center}  \fontsize{8pt}{10pt}\selectfont
   \caption{  Summary of  the (rounded) Second digit ($Sd$)  of the downloaded   $PI$, further  distinguishing between $PI^{(-)}$	 and $PI^{(+)}$,  and the pertinent frequencies to be compared to the BL2.} \label{tablePIBL2}
\begin{tabular}{|c||c|c|c||c|c|c||c|c|c|} 
 \hline 
$d_2$ &$Sd$ $ PI$&$Sd$  $PI^{(-)}$& $Sd$  $PI^{(+)}$&   BL2 $PI$& BL2   $PI^{(-)}$ &   BL2 $PI^{(+)}$  & BL2  \\\hline 
0	&	875	&	166	&	709	&	0.12928	&	0.14310	&	0.12643	&	0.11968	\\
1	&	753	&	121	&	632	&	0.11126	&	0.10431	&	0.11270	&	0.11389	\\
2	&	745	&	129	&	616	&	0.11008	&	0.11121	&	0.10984	&	0.10882	\\
3	&	707	&	126	&	581	&	0.10446	&	0.10862	&	0.10360	&	0.10433	\\
4	&	675	&	116	&	559	&	0.09973	&	0.10000	&	0.09968	&	0.10031	\\
5	&	626	&	118	&	508	&	0.09249	&	0.10172	&	0.09058	&	0.09668	\\
6	&	648	&	94	&	554	&	0.09574	&	0.08103	&	0.09879	&	0.09337	\\
7	&	606	&	115	&	491	&	0.08954	&	0.09914	&	0.08755	&	0.09035	\\
8	&	595	&	78	&	517	&	0.08791	&	0.06724	&	0.09219	&	0.08757	\\
9	&	538	&	97	&	441	&	0.07949	&	0.08362	&	0.07864	&	0.08500	\\\hline \hline
$\chi^2$ &9.855&15.208 &10.742&&&&\\
$MAD$ &&&&0.02741&0.08787&0.03560&\\
	\hline
	 &\multicolumn{3}{|c||}{$\chi^2_c$ =   16.919,  at  5\%,\; for D = 9. } &\multicolumn{3}{|c||}{close conformity if MAD $\leq$ 0.008}  & \\	\hline
	\end{tabular}
 \end{center}
 \end{table} 

   \begin{table}\begin{center}  \fontsize{8pt}{10pt}\selectfont
   \caption{  Summary of  the (rounded) Second digit ($Sd$)  of the downloaded  $TA$ and $PI/TA$ further  distinguishing  and their pertinent frequencies to be compared to the BL2.} \label{tableTAPITABL2}
\begin{tabular}{|c||c|c||c|c||c|c|c|} 
 \hline 
$d_2$ &$Sd$ $ TA$&$Sd$  $PI/TA$ &   BL2 $TA$& BL2   $PI/TA$ & BL2  \\\hline 
0	&	821	&	853	&	0.12054	&	0.12609	&	0.11968	\\
1	&	764	&	854	&	0.11217	&	0.12624	&	0.11389	\\
2	&	758	&	696	&	0.11129	&	0.10288	&	0.10882	\\
3	&	717	&	681	&	0.10527	&	0.10067	&	0.10433	\\
4	&	686	&	666	&	0.10072	&	0.09845	&	0.10031	\\
5	&	635	&	650	&	0.09323	&	0.09608	&	0.09668	\\
6	&	621	&	594	&	0.09118	&	0.08780	&	0.09337	\\
7	&	634	&	605	&	0.09308	&	0.08943	&	0.09035	\\
8	&	616	&	574	&	0.09044	&	0.08485	&	0.08757	\\
9	&	559	&	592	&	0.08207	&	0.08751	&	0.08500	\\	\hline\hline
$\chi^2$ &3.6678&18.063&&&\\
$MAD$ &&&0.02057&0.04253&\\
	\hline
	 &\multicolumn{2}{|c||}{$\chi^2_c$ =   16.919, at  5\%,\; for D = 9 } &\multicolumn{2}{|c||}{close conformity if MAD $\leq$ 0.008}  & \\	\hline
	\end{tabular}
 \end{center}
 \end{table}

   \begin{table}\begin{center}  \fontsize{8pt}{10pt}\selectfont
   \caption{  Summary of  the (rounded)  First- and Second  digit ($FSd$) of the downloaded   $PI$, further  distinguishing between $PI^{(-)}$	 and $PI^{(+)}$,  and the pertinent frequencies to be compared to the BL12.} \label{tablePIBL12}
\begin{tabular}{|c||c|c|c||c|c|c||c|c|c|} 
 \hline 
$d_{12}$ &$FSd$ $ PI$&$FSd$  $PI^{(-)}$& $FSd$  $PI^{(+)}$&   BL12 $PI$& BL12   $PI^{(-)}$ &   BL12 $PI^{(+)}$  & BL12  \\\hline 
10	&	264	&	56	&	208	&	0.039007	&	0.048276	&	0.037090	&	0.04139	\\
11	&	251	&	43	&	208	&	0.037086	&	0.037069	&	0.037090	&	0.03779	\\
12	&	246	&	46	&	200	&	0.036348	&	0.039655	&	0.035663	&	0.03476	\\
 ... 	&	 ...	&	 ...	&	 ...	&	... 		&	 ...		&		...	&	...		\\
20	&	162	&	31	&	131	&	0.023936	&	0.026724	&	0.023359	&	0.02119	\\
21	&	148	&	21	&	127	&	0.021868	&	0.018103	&	0.022646	&	0.02020	\\
 ... 	&	 ...	&	 ...	&	 ...	&	... 		&	 ...		&		...	&	...		\\
30	&	100	&	18	&	82	&	0.014775	&	0.015517	&	0.014622	&	0.01424	\\
  ... 	&	 ...	&	 ...	&	 ...	&	... 		&	 ...		&		...	&	...		\\
 50	&	56	&	11	&	45	&	0.008274	&	0.009483	&	0.008024	&	0.00860\\
  ... 	&	 ...	&	 ...	&	 ...	&	... 		&	 ...		&		...	&	...		\\
 90	&	39	&	9	&	30	&	0.005762	&	0.007759	&	0.005350	&	0.00480	\\
 ... 	&	 ...	&	 ...	&	 ...	&	... 		&	 ...		&		...	&	...		\\
99	&	19	&	4	&	15	&	0.002807	&	0.003448	&	0.002675	&	0.00436	\\
	\hline
$\chi^2$ &116.213&93.989&128.654&&&&\\
$MAD$ &&&&0.09418&0.21754&0.11131&\\
	\hline
	 &\multicolumn{3}{|c||}{$\chi^2_c$ = 113.145, at  5\%,\; for D = 89} &\multicolumn{3}{|c||}{close conformity if MAD $\leq$ 0.0012}  & \\	\hline
	\end{tabular}
 \end{center}
 \end{table} 

   \begin{table}\begin{center}  \fontsize{8pt}{10pt}\selectfont
   \caption{ Summary of  a few values of the (rounded) First- and Second-  digit ($FSd$)  of the downloaded   $TA$ and $PI/TA$ data and the pertinent frequencies to be compared to the BL12.} \label{tableTAPITABL12}
\begin{tabular}{|c||c|c||c|c||c|} 
 \hline 
$d_{12}$ &$FSd$ $ TA$&$FSd$  $PI/TA$&     BL12 $TA$& BL12   $TA/PI$   & BL12  \\\hline 

10	&	 308	&	343 	&	0.04522	&	0.05070	&	 0.04139	\\
11	&	 259	&	318	&	0.03803	&	0.04701	&	0.03779	\\
12	&	 233	&	249	&	0.03421	&	0.03681	&	0.03476	\\
... 	&	 ...	&	 ...	&	 ...		&	... 		&	 ...		\\
20	&	144	&	154	&	0.02114	&	0.02276	&	0.02119	\\
21	&	135	&	150	&	0.01982	&	0.02217	&	0.02020	\\
 ... 	&	 ...	&	 ...	&	 ...		&	 ...		&	 ...		\\
 30 	&	 95	&	74	&	0.01395	&	0.01094	&	0.01424	\\
  ... 	&	 ...	&	 ,,,	&	 ...		&	 ...		&	 ...		\\
 50	&	 63	&	44	&	0.00925	&	0.00650	&	0.00860\\
 ... 	&	 ...	&	 ...	&	 ...		&	 ...		&	 ...		\\
 90	&	 32	&	30	&	0.00470	&	0.00443	&	0.00480	\\
 ... 	&	 ...	&	 ...	&	 ...		&	 ...		&	 ...		\\
99	&	19 	&	38 	&	 0.00279	&	0.00562	&	0.00436	\\
\hline
$\chi^2$ &26.5234&196.574&&&\\
$MAD$ &&&0.04658&0.13625&\\
	\hline
	 &\multicolumn{2}{|c||}{$\chi^2_c$ =  113.145, at  5\%,\; for D = 89 } &\multicolumn{2}{|c||}{close conformity if MAD $\leq$ 0.0012}  & \\	\hline
	\end{tabular}
 \end{center}
 \end{table} 

\section{Conclusions} \label{conclusions}
In brief, recall that we aim to assess the  discrepancy between  secondary (financial) data and  their possible theoretical Benford distributions, through two statistical tests based on different concepts: the $\chi^2$ test and  the Mean Absolute Deviation (MAD), both outlined  in  Section \ref{findings}. 
We have used rather large financial data, i.e., the pre-tax income ($PI$)  and the total assets ($TA$) of 567 companies listed on the FTSE All-Share index, gathered from the Refinitiv EIKON database, covering 14 years, i.e., the period from 2009 to 2022, as described in Section \ref{dataandmethodology}. Since the validity and applicability of Benford's laws (BLs) is still debated, in particular on indirect measures, we have also derived the $PI/TA$  data set. Furthermore, due to the data size (a little bit less than 7000 data points), we have examined   cases of  either positive or negative $PI$.

Indeed, Benford's Law, which describes the frequency distribution of digits in many real-world datasets, has been extensively applied in accounting to detect anomalies such as rounding up or other irregularities. While much of the research focuses on the first digit, some studies (including the present study) have also examined the distribution of second digits and the first and second digits' occurrence 
on income and total assets  data integrity.
Benford's Law provides a method to detect such anomalies by comparing the expected distribution of digits to the observed distribution in the dataset.
The present study has examined such deviations for first two digits for the $PI$ variable in the profits and losses domain, identifying significant deviations for $PI^(+)$ and $PI^(-)$ from the theoretical distribution.

  {\color{black} 
 Thus, whether 
the actual $TA$, $PI$, $PI/TA$ proportions do not statistically differ from the proportion expected from Benford’s Laws (BL2 and BL12) according to the $\chi^2$ and MAD statistical tests, i.e., the null hypothesis, can be now verified.
}

From the Tables summarizing the MAD values, it seems obvious that all data  appear to be non-conforming to BLs.   {\color{black} This null hypothesis  is rejected.}

In contrast, conclusions from the $\chi^2$ tests are more ambiguous. Indeed the smallest $\chi^2$  with a value below the critical one for the pertinent number of degrees of freedom occur for several cases:

firstly, for $Fd$ $PI^{(+)}$, 

secondly, for $Sd$ $PI$,  $Sd$ $PI^{(-)}$, $SDd$ $PI^{(+)}$, and 
$Sd $ $TA$,

and finally, for $FSd$ $PI^{(-)}$ and $FSd$ $TA$.

A few cases are close to the $\chi^2$ critical value like

$Fd$ $ PI $ and $Fd$ $PI^{(+)}$,  

$Sd$ $ TA$,

and $FSd$ $ PI$,  and  $FSd$ $PI^{(+)}$.

It is remarkable that no $PI/TA$ ($Fd$, $Sd$, and $FSd$) ratio obeys the 5\% $\chi^2$-test criterion. Although there is no proof of the following deduction, one may assume that  this finding might be due to the bizarre $TA/PI$ distribution, - illustrated on Fig. \ref{figPlot4TAPI}.   

In fact, this observation  allows us to point to one still opened question on the applicability, and further validity, of Benford's laws for derived measures, if it is found that the  initial, raw, measures is found to obey the empirical BLs. Our results through the $\chi^2$-test values seem to indicate that the universally extended validity is  dubious.  {\color{black} As a suggestion for further work  the $U^2$ test would seem to be seriously considered \cite{Lesperanceetal16,Ducharmeetalarxiv}.}

In conclusion, one can again be amazed that two different statistical tests do not lead to similar deductions on the conformity of large, {\it a priori} unmanipulated, data sets. On the other hand, one may further question the lack of no conformity with respect to the MAD test. Indeed, this implies that the data might be manipulated, likely in order to maintain a balance between the reported income, thus lowering the tax amount,  and the positive show of benefits for shareholders. Nevertheless, our findings add to the observations of Alali and Romero \cite{Alali and Romero 2013}, on one hand, and on those of Henselmann et al.  \cite{Henselmann et al. 2015} on the other hand. Although we have used quite different data sets, and different tests, we confirm that much  remains to be understood and explained.

To sum up, this note enriches the application of Benford's Law as a robust tool for detecting anomalies and potential fraud in accounting data, utilizing an extensive dataset derived from FTSE-listed companies. The Note also stresses the need for more elaborate statistical analyses before drawing conclusions.

 {\color{black}   Thus, future research directions may be here highlighted, in particular studies on other financial data, through BLs,  - if one is interested in financial data manipulation, but also through other statistical tests, thereby agreeing with  recent  recommendations by Lesperance et al. \cite{Lesperanceetal16}, Ducharme et al. \cite{Ducharmeetalarxiv}, Henselmann et al.   \cite{Henselmann et al. 2015}, 
  Cerqueti and Lupi \cite{CerquetiLupi2023,CerquetiLupi2021},  and  Barabesi et al. \cite{Barabesietal21}.}

\vspace{6pt}

{\bf Authors Contributions}: Conceptualization, M.A., P.E.S., and P.K.; methodology, M.A., P.E.S., and P.K.; software, M.A. and P.E.S.; validation, M.A., P.E.S. and P.K.; formal analysis, M.A. and P.E.S.; investigation, M.A.; resources, P.E.S.; data curation, M.A. and P.E.S.; writing---original draft preparation, M.A.; writing---review and editing, M.A., P.E.S., and P.K.; visualization, M.A. All authors have read and agreed to the published version of the manuscript.

{\bf Funding}: MA was partially supported by the project ‘’A better understanding of socio-economic systems using quantitative methods from physics’’'   funded by European Union---NextgenerationEU and Romanian Government, under National Recovery and Resilience Plan for Romania, contract no.760034/23.05.2023, code PNRR-C9-I8-CF 255/29.11.2022, through the Romanian Ministry of Research, Innovation and Digitalization, within Component 9, ``Investment i8''.



{\bf Data availability}:   Details regarding from where the analyzed  data  can be obtained is included in the text.
It is also explained how new data was created.  The final numerical results are all reported. 


{\bf Acknowledgements}:   {\color{black} 
We acknowledge comments and suggestions by anonymous reviewers.  }

{\bf Conflicts of interest}: The authors declare no conflicts of interest.
The  MA's funders had no role in the design of the study, in the collection, analyses, or interpretation of data, in the writing of the manuscript, or in the decision to publish the results.

\end{document}